\documentclass[aps,preprint]{revtex4}%
\usepackage{amsfonts}
\usepackage{amsmath}
\usepackage{amssymb}
\usepackage{graphicx}
\usepackage{pdfpages}
\usepackage{cleveref}
\usepackage{subcaption}%
\providecommand{\U}[1]{\protect\rule{.1in}{.1in}}
\providecommand{\U}[1]{\protect\rule{.1in}{.1in}}

\begin{document}
\preprint{HEP/123-qed}
\title{GUP Modified Hawking Radiation in Bumblebee Gravity}
\author{Sara Kanzi}
\affiliation{}
\author{\.{I}zzet Sakall{\i}}
\affiliation{Physics Department, Arts and Sciences Faculty, Eastern Mediterranean
University, Famagusta, North Cyprus via Mersin 10, Turkey.}
\author{}
\affiliation{}

\begin{abstract}
The effect of Lorentz symmetry breaking (LSB) on the Hawking radiation of
Schwarzschild-like black hole found in the bumblebee gravity model (SBHBGM) is
studied in the framework of quantum gravity. To this end, we consider Hawking
radiation spin-0 (bosons) and spin-$\frac{1}{2}$ particles (fermions), which
go in and out through the event horizon of the SBHBGM. We use the modified
Klein-Gordon and Dirac equations, which are obtained from the generalized
uncertainty principle (GUP) to show how Hawking radiation is affected by the
GUP and LSB. In particular, we reveal that, independent of the spin of the
emitted particle, GUP causes a change in the Hawking temperature of the
SBHBGM. Furthermore, we compute the semi-analytic greybody factors (for both
bosons and fermions) of the SBHBGM. Thus, we reveal that LSB is effective on
the greybody factor of the SBHBGM such that its redundancy decreases the value
of the greybody factor. Our findings are graphically depicted.

\end{abstract}
\volumeyear{ }
\eid{ }
\date[09.03.2018]{}
\received{}

\maketitle
\tableofcontents

\section{INTRODUCTION}

In spite of their overwhelming successes in describing nature, General
Relativity (GR) (i.e., detection of the gravitational waves
\cite{Abbott2016v1,Abbott2016v2} and observation of the shadow of the M87
supermassive black hole (BH) \cite{Akiyama2019}) and Standard Model (SM)
(i.e., detection of the Higgs boson \cite{HiggsBoson}) of particle physics are
incomplete theories. While Einstein's theory of GR successfully describes
gravity at a classical level, SM explains particles and the other three
fundamental forces (electromagnetic, and the strong and weak nuclear forces)
at a quantum level. The unification of GR and SM is a fundamental quest, and
this success will necessarily lead us to a deeper understanding of nature. In
the search for this unification, some quantum gravity theories (QGTs) have
been proposed, but direct tests of their features are beyond the energy scale
of the currently available experiments. Because, they will be observed on the
Planck scale which is around $10^{19}$ ($GeV$). However, it is possible that
some signals of the QGT appear at sufficiently low energy scales and their
effects can be observed in experiments on existing energy scales. One of these
signals could be related to the LSB \cite{Nomura2013}.

The theory of LSB has been under intense research since the proposed SM
Extension (SME)
\cite{SME1v1,SME1v2,SME2,SME3v1,SME3v2,SME3v3,SME3v4,SME4v1,SME4v2,SME4v3},
which is an effective field theory that includes the SM, GR, and every
possible operator that breaks the Lorentz symmetry. With the SME, further
investigations of the LSB can be made in the context of high energy particle
physics, nuclear physics, gravitational physics, and astrophysics. The
simplest models that contain a vector field which dynamically breaks the
Lorentz symmetry are called bumblebee models
\cite{mybumblev1,mybumblev2,mybumblev3,mybumblev4,mybumblev5}.These models,
although owning a simpler form, have interesting features such as rotations,
boosts, and CPT violations. In a bumblebee gravity model (BGM), potential $V$
is included in the action $S_{BM}$, which evokes a vacuum expectation value
(VEV) for the vector field. The potential $V$ is formed as a function of a
scalar combination $\aleph$ of the vector $B_{\mu}$ and the metric $g_{\mu\nu
}$ (plus the other matter fields, if there are any). The potential has a
minimum at $\frac{dV}{d\aleph}=0$. At the $V_{\min}$, the bumblebee field
$B_{\mu}$ incorporates a vacuum value shown by $\left\langle B_{\mu
}\right\rangle =b_{\mu}$, which is the so-called vacuum vector. In fact, the
vacuum vector is nothing but a background vector that gives rise to local
(spontaneous) LSB \cite{myLSB}. The scalar of the BGM, in general, reads as
$\aleph=\left(  B^{\mu}B_{\mu}\pm b^{2}\right)  $ in which $b$ is a constant
having dimensions of mass ($M$). Thus, the $V_{\min}$ satisfies the condition
of $\frac{dV}{d\aleph}=0$ for $\aleph=0$. Here, $b_{\mu}$ is spontaneously
induced as a timelike vector abiding by $b_{\mu}b^{\mu}=-b^{2}$. For instance,
the aether models \cite{aether1,aether1v,aether2,aether3}\ are based on a
vector field, which is in the Lagrangian density of the system with a
non-vanishing VEV. The vector field dynamically selects a preferred frame at
each point in the considered spacetime and spontaneously breaks the Lorentz
invariance. This is a mechanism reminiscent of the breaking of local gauge
symmetry described by the Higgs mechanism. In general, the subclass of aether
models obeys the following action \cite{aetherAct}:%

\begin{equation}
S_{BM}=\int d^{4}x\left[  \frac{1}{16\pi G}\left(  R+%
\hbox{\rm\rlap/c}%
B^{\mu}B^{\nu}R_{\mu\nu}\right)  -\frac{1}{4}B^{\mu\nu}B_{\mu\nu}%
-V\aleph\right]  , \label{NN1}%
\end{equation}

where the parameter $%
\hbox{\rm\rlap/c}%
,$ having dimensions of $M^{-2},$ denotes the coupling between the Ricci
tensor ($R_{\mu\nu}$) and $B^{\mu}$. $B_{\mu\nu}$ is the bumblebee field
strength:
\begin{equation}
B_{\mu\nu}=\nabla_{\mu}B_{\nu}-\nabla_{\nu}B_{\mu}. \label{NN2}%
\end{equation}
As mentioned above, $V$ is the potential of the bumblebee field that drives
the breaking of the Lorentz symmetry of the Lagrangian by collapsing onto a
non-zero minimum at $\aleph=0$\ or $B_{\mu}B^{\mu}=\mp b^{2}.$ In fact,
$B_{\mu}$ is one of the Lorentz breaking coefficients and it shows a preferred
direction in which the equivalence-principle is locally broken for a certain
Lorentz frame. Observations of Lorentz violation can emerge if the particles
or fields interact with the bumblebee field \cite{aetherAct}. It is worth
noting that when a smooth quadratic potential is chosen as
\begin{equation}
V=A\aleph^{2}, \label{NN3}%
\end{equation}
where $A$ is a dimensionless constant, one gets the Nambu-Goldstone
excitations (massless bosons) besides the massive excitations \cite{aetherAct}%
. Besides, the linear Lagrange-multiplier potential is given by $V=\lambda
\aleph.$These potentials (\ref{NN1})\ and (\ref{NN2}) present also the
breaking of the $U(1)$ gauge invariance and other implications to the behavior
of the matter sector, the photon, and the graviton. For a topical review (from
experimental proposals to the test results) of the BGMs, the reader is
referred to \cite{mybumblev4} and references therein. Furthermore, the studies
using the bumblebee models have gained momentum for the last two decades. The
vacuum solutions for the bumblebee field for purely radial, temporal-radial,
and temporal-axial Lorentz symmetry breaking were obtained in \cite{myex1}.
New spherically static black hole (BH) \cite{myex2} and traversable wormhole
\cite{myex3} solutions in the BGM have been recently discovered. Bluhm
\cite{myex4} discussed the Higgs mechanism in the BGM. The electrodynamics of
the bumblebee fields was studied by \cite{myex5} in which the bumblebee field
was considered as a photon field. Propagation velocity of the photon field,
along with its possible effects on the accelerator physics and cosmic ray
observations, was also investigated. BGMs are also used to limit the
likelihood of Lorentz violation in astrophysical objects such as the Sun
\cite{myex6}. For other studies demonstrating the physical effects
(quasinormal modes, thermodynamics, etc.) of the bumblebee field, the reader
may refer to
\cite{mybumble2v1,mybumble2v2,mybumble2v3,mybumble2v4,mybumble2v5,mybumble2v6,mybumble2v7,mybumble2v8,mybumble2v9,mybumble2v10}
and references therein.

Hawking's ground-breaking studies \cite{myHR1,myHR2} can be considered as the
onset of QGT \cite{myWald,WaldGBF}. Since then there have been numerous
research papers on the subject of Hawking radiation (HR) in the literature
(see, for instance,
\cite{5v1,5v2,5v3,5v4,5v5,5v6,5v7,5v8,5v9,5v10,5v11,5v12,5v13,5v14,5v15,5v16,5v17,5v18,5v19,5v20,5v21,6v1,6v2,6v3,6v4,6v5,6v6,6v7,6v8,6v9,6v10,6v11,6v12,6v13}%
). Several methods have been developed to calculate the HR of BHs
\cite{7,8,9,10}. In this study, we mainly focus on the quantum gravity effects
on the HR of SBHBGM \cite{myex2} in the tunneling paradigm. Although a number
of QGTs have been proposed, however, physics literature does not as yet have a
complete and consistent QGT. In the absence of a complete quantum description
of the HR, we use effective models to describe the quantum gravitational
behavior of the BH evaporation. In particular, string theory, loop quantum
gravity, and quantum geometry predict the minimal observable length on the
Planck scale \cite{11,12}, which leads to the GUP
\cite{GUP13v1,GUP13v2,GUP13v3,GUP14v1,GUP14v2,GUP14v3,GUP14v4,GUP14v5,GUP14v6,GUP14v7,GUP14v8,GUP14v9,GUP14v10,GUP14v11,GUP14v12,GUP14v13,GUP14v14,GUP14v15,GUP14v16,GUP14v17,GUP14v18,GUP14v19,GUP14v20,GUP14v21,GUP14v22,GUP15v1,GUP15v2,GUP15v3,GUP15v4,GUP15v5,GUP15v6,GUP15v7,GUP15v8,GUP15v9,GUP15v10,GUP15v11,GUP15v12}%
:
\begin{equation}
\Delta x\Delta p\geq\frac{\hbar}{2}\left[  1+\beta(\Delta p)^{2}\right]  ,
\label{maq}%
\end{equation}
where $\beta=\frac{\alpha_{0}}{M_{p}^{2}}$ in which $M_{p}=\sqrt{\frac{\hbar
c}{G}}$ denotes the Planck mass and $\alpha_{0}$ is the dimensionless
parameter, which encodes the quantum gravity effects on the particle dynamics.
The upper bound for $\alpha_{0}$ was obtained as $\alpha_{0}<10^{21}$
\cite{16}. Today, the effects of GUP on BHs have been extensively studied in
the literature \cite{17,18,19}. To amalgamate the GUP with the considered wave
equation, the Wentzel-Kramers-Brillouin approximation \cite{20} is generally
used. Thus, one can obtain the quantum corrections to the HR of the BH
\cite{21,22}.

Since the GUP and LSB effects are high energy modifications of the QGT, it is
interesting to investigate their combined effects. To this end, we study the
GUP-assisted HR of bosons (spin-$0$) and fermions' (spin-$\frac{1}{2}$)
tunneling \cite{24,25} from the SBHBGM. Although the SBHBGM looks like the
Schwarzschild BH, the differences in the Kretschmann scalars confirm that both
BHs are physically different. The effects of spin and Lorentz-violating
parameter $L$ \cite{26,27}\ on the quantum corrected HR are analyzed. We also
study the problem of low energy greybody factors \cite{28,29} for the bosons
and fermions emitted by the SBHBGM. For this purpose, we implement a method
developed by Unruh \cite{30,31}. It is also worth noting that Lorentz
invariant massive gravity can be obtained dynamically from spontaneous
symmetry breaking in a topological Poincare gauge theory \cite{massive1}.
Besides, BH radiation in massive gravity (selecting a preferred direction of
time) naturally corresponds to violations of the Lorentz symmetry
\cite{massive2v1,massive2v2,massive2v3}.

The outline of the paper is as follows: In Sec. 2, we briefly introduce the
SBHBGM and discuss some of its basic features. Section 3 is devoted to the
computation of GUP-corrected HR of the bosons' tunneling from the SBHBGM. In
Sec. 4, we compute the quantum tunneling rate for the fermions of the SBHBGM
using the GUP-modified Dirac equation and derive the modified HR. In the
following section, we derive the greybody factor of the SBHBGM. In Sec. 6, we
summarize our results. (Throughout the paper, we use geometrized units:
$c=G=1$.)

\section{SBHBGM SPACETIME}

The Lagrangian density of the BGM \cite{32,33} yields the following extended
vacuum Einstein equations
\begin{equation}
G_{\mu\nu}=R_{\mu\nu}-\frac{1}{2}Rg_{\mu\nu}=\kappa T_{\mu\nu}^{B}\text{,}
\label{iz1}%
\end{equation}
where $G_{\mu\nu}$ and $T_{\mu\nu}^{B}$ {are the Einstein and bumblebee
energy-momentum tensors, respectively. }${\kappa=8\pi G}_{{N}}$ is the
gravitational coupling and {$T_{\mu\nu}^{B}$ is given by}
\begin{align}
T_{\mu\nu}^{B}  &  =-B_{\mu\alpha}B_{~\nu}^{\alpha}-\frac{1}{4}B_{\alpha\beta
}B^{\alpha\beta}g_{\mu\nu}-Vg_{\mu\nu}+2V^{\prime}B_{\mu}B_{\nu}+\frac{\xi
}{\kappa}\left[  \frac{1}{2}B^{\alpha}B^{\beta}R_{\alpha\beta}g_{\mu\nu
}-B_{\mu}B^{\alpha}R_{\alpha\nu}\right. \nonumber\\[0.08in]
&  \left.  -B_{\nu}B^{\alpha}R_{\alpha\mu}+\frac{1}{2}\nabla_{\alpha}%
\nabla_{\mu}\left(  B^{\alpha}B_{\nu}\right)  +\frac{1}{2}\nabla_{\alpha
}\nabla_{\nu}\left(  B^{\alpha}B_{\mu}\right)  -\frac{1}{2}\nabla^{2}\left(
B_{\mu}B_{\nu}\right)  -\frac{1}{2}g_{\mu\nu}\nabla_{\alpha}\nabla_{\beta
}\left(  B^{\alpha}B^{\beta}\right)  \right]  , \label{iz2}%
\end{align}
where $\xi$ is the real coupling constant (having dimension $M^{-1}$) that
controls the non-minimal gravity-bumblebee interaction. From now on,
\textit{the prime symbol shall denote the differentiation with respect to its
argument}. Meanwhile, there are other generic bumblebee models having non-zero
torsion in the literature (see for instance \cite{32}). In Eq. (\ref{iz2}),
the potential $V\equiv V\left(  \aleph\right)  $ provides a non-vanishing VEV
for $B_{\mu}$. As it was stated above (see also \cite{36,37}), the VEV of the
bumblebee field is determined when $V=V^{\prime}=0.$ Taking the covariant
divergence of the bumblebee Einstein equations (\ref{iz1}) and using the
contracted Bianchi identities, one gets
\begin{equation}
\nabla^{\mu}T_{\mu\nu}^{B}=0\text{,} \label{iz4}%
\end{equation}
which gives the covariant conservation law for the {bumblebee total}
energy-momentum tensor $T_{\mu\nu}$. {Thus, Eq. (\ref{iz1}) reduces to}
\begin{equation}
R_{\mu\nu}=\kappa T_{\mu\nu}^{B}+\frac{\xi}{4}g_{\mu\nu}\nabla^{2}\left(
B_{\alpha}B^{\alpha}\right)  +\frac{\xi}{2}g_{\mu\nu}\nabla_{\alpha}%
\nabla_{\beta}(B^{\alpha}B^{\beta})\text{.} \label{iz5}%
\end{equation}
One can immediately see that when the bumblebee field $B_{\mu}$ vanishes, we
recover the ordinary Einstein equations. Recently, {the vacuum solution in the
BGM induced by the LSB has been derived by Casana \textit{et al.}
\cite{myex2}. The solution is obtained when the bumblebee field }$B_{\mu}$%
{{}remains frozen in its VEV }$b_{\mu}$ \cite{38,38n}{. Namely, we have }%
\begin{equation}
B_{\mu}=b_{\mu}\text{, \ \ \ }\Rightarrow\text{\ \ \ \ \ \ \ }{b}%
\text{{$_{\mu\nu}\equiv\partial_{\mu}$}}{b}\text{{$_{\nu}-\partial_{\nu}$}}%
{b}\text{{$_{\mu}$}.} \label{iz6}%
\end{equation}

{Thus, the extended Einstein equations are found to be }%
\begin{align}
&  R_{\mu\nu}+\kappa b_{\mu\alpha}b_{~\,\nu}^{\alpha}+\frac{\kappa}%
{4}b_{\alpha\beta}b^{\alpha\beta}g_{\mu\nu}+\xi b_{\mu}b^{\alpha}R_{\alpha\nu
}+\xi b_{\nu}b^{\alpha}R_{\alpha\mu}-\frac{\xi}{2}b^{\alpha}b^{\beta}%
R_{\alpha\beta}g_{\mu\nu}-\nonumber\\[0.08in]
&  \frac{\xi}{2}\nabla_{\alpha}\nabla_{\mu}\left(  b^{\alpha}b_{\nu}\right)
-\frac{\xi}{2}\nabla_{\alpha}\nabla_{\nu}\left(  b^{\alpha}b_{\mu}\right)
+\frac{\xi}{2}\nabla^{2}\left(  b_{\mu}b_{\nu}\right)  =0\text{.} \label{iz7}%
\end{align}

Assuming {a spacelike background for }$b_{\mu}${ as}
\begin{equation}
b_{\mu}=[0,b_{r}(r),0,0]\text{,} \label{iz8}%
\end{equation}

and using the condition $b^{\mu}b_{\mu}=b^{2}=$\textit{constant}, LSB
parameter ($L$) is defined as $L=\xi b^{2}\geq0$ \cite{32}. A spherically
symmetric static vacuum solution to Eq. (\ref{iz7}) is obtained as follows
\cite{myex2}{ }
\begin{equation}
ds^{2}=-\left(  1-\frac{2M}{r}\right)  dt^{2}+\left(  1+L\right)  \left(
1-\frac{2M}{r}\right)  ^{-1}dr^{2}+r^{2}\left(  d\theta^{2}+\sin^{2}\theta
d\varphi^{2}\right)  \text{,} \label{iz9}%
\end{equation}

which we call it SBHBGM solution. This BH solution represents a purely radial
Lorentz-violation outside a spherical body characterizing a modified BH
solution. In the limit $L\rightarrow0~(b^{2}\rightarrow0),$ one can
immediately see that the usual Schwarzschild metric is recovered. For the
metric (\ref{iz9}), the Kretschmann scalar becomes
\begin{equation}
\mathcal{K}=\frac{4\left(  12M^{2}+4LMr+L^{2}r^{2}\right)  }{r^{6}\left(
1+L\right)  ^{2}}\text{,} \label{iz10}%
\end{equation}
{{}}

which is different than the Kretschmann scalar of the Schwarzschild BH. It
means that none of the coordinate transformations link the metric (\ref{iz9})
to the usual Schwarzschild BH. When $r=2M$, Eq. (\ref{iz9}) becomes finite:
the coordinate singularity can be removed by applying a proper coordinate
transformation. However, in the case of $r=0$, physical singularity cannot be
removed. So, we see that the behaviors of the physical ($r=0$) and coordinate
($r=r_{h}=2M$ : event horizon) singularities do not change in the BGM.

The Hawking temperature of the metric (\ref{iz9}) can be computed from Eq.
(1), in which the surface gravity is given by \cite{myWald}%

\begin{equation}
\kappa=\nabla_{\mu}\chi^{\mu}\nabla_{\nu}\chi^{\nu}, \label{iz12}%
\end{equation}

where $\chi^{\mu}$ is the timelike Killing vector field. Thus, the Hawking
temperature of the SBHBGM (\ref{iz9}) reads%

\begin{equation}
T_{H}=\frac{1}{4\pi\sqrt{-g_{tt}g_{rr}}}\left.  \frac{dg_{tt}}{dr}\right\vert
_{r=r_{h}}=\left.  \frac{1}{2\pi\sqrt{1+L}}\frac{M}{r^{2}}\right\vert
_{r=r_{h}}=\frac{1}{8\pi M\sqrt{1+L}}. \label{iz13}%
\end{equation}

One can easily see from Eq. (\ref{iz13}) that the non-zero LSB parameter has
the effect of reducing the Hawking temperature of a Schwarzschild BH.

\section{GUP ASSISTED HR OF SBHBGM: BOSONS' TUNNELING}

The generic Klein-Gordon equation within the framework of GUP is given by
\cite{36}%

\begin{equation}
-(i\hbar)^{2}\partial^{t}\partial_{t}\Psi=\left[  (i\hbar)^{2}\partial^{\mu
}\partial_{\mu}+m^{2}\right]  \left\{  1-2\beta\left[  (i\hbar)^{2}%
\partial^{\mu}\partial_{\mu}+m^{2}\right]  \right\}  \Psi, \label{iz14}%
\end{equation}

where $\beta$ and $m$ are the GUP parameter and mass of the scalar particle,
respectively. Introducing the following ansatz for the wave function $\Psi$
\begin{equation}
\Psi=\exp\left[  \frac{i}{\hbar}I(t,r,\theta,\varphi)\right]  , \label{iz15}%
\end{equation}

where $I(t,r,\theta,\varphi)$ is the classically forbidden action for quantum
tunneling. Substituting Eq. (\ref{iz15}),\ together with the metric functions
of line-element (\ref{iz9}), into Eq. (\ref{iz14}), we get%
\[
(f)^{-1}(\partial_{t}I)^{2}=\left[  \frac{f}{1+L}(\partial_{r}I)^{2}+\frac
{1}{r_{h}^{2}}(\partial_{\theta}I)^{2}+\frac{1}{r_{h}^{2}\sin^{2}\theta
}(\partial_{\varphi}I)^{2}+m^{2}\right]
\]

\begin{equation}
\times\left\{  1-2\beta\left[  \frac{f}{1+L}(\partial_{r}I)^{2}+\frac{1}%
{r^{2}}(\partial_{\theta}I)^{2}+\frac{1}{r^{2}\sin^{2}\theta}(\partial
_{\varphi}I)^{2}+m^{2}\right]  \right\}  , \label{iz16}%
\end{equation}

where%

\begin{equation}
f=1-\frac{2M}{r}. \label{iz17}%
\end{equation}

It is easy to see that SBHBGM (\ref{iz9}) admits two Killing vectors
$<\partial_{t},\partial_{\varphi}>$. The existence of these symmetries implies
that we can assume a following separable solution for the action%
\begin{equation}
I=-\omega t+R(r)+S(\theta)+J\varphi, \label{iz18}%
\end{equation}

where $\omega$ and $J$ denote the energy and angular momentum of the radiated
particle, respectively. Substituting Eq. (\ref{iz18}) in Eq. (\ref{iz16}), we
obtain
\[
\frac{\omega^{2}}{f}=\left[  \frac{f}{1+L}(\partial_{r}R)^{2}+\frac{1}{r^{2}%
}\left(  (\partial_{\theta}S)^{2}+\frac{J^{2}}{\sin^{2}\theta}\right)
+m^{2}\right]
\]

\begin{equation}
\times\left\{  1-2\beta\left[  \frac{f}{1+L}(\partial_{r}R)^{2}+\frac{1}%
{r^{2}}\left(  (\partial_{\theta}S)^{2}+\frac{J^{2}}{\sin^{2}\theta}\right)
+m^{2}\right]  \right\}  . \label{iz19}%
\end{equation}

We focus only on the radial trajectories in which only the $(r-t)$ sector is
considered. Thus, one can set%

\begin{equation}
\frac{1}{r^{2}}\left(  (\partial_{\theta}S)^{2}+\frac{J^{2}}{\sin^{2}\theta
}\right)  =e, \label{iz20}%
\end{equation}

where $e$ is a constant. So, Eq. (\ref{iz19}) becomes%
\begin{equation}
\left[  \frac{f}{1+L}(\partial_{r}R)^{2}+e+m^{2}\right]  \left\{
1-2\beta\left[  \frac{f}{1+L}(\partial_{r}R)^{2}+e+m^{2}\right]  \right\}
=\frac{\omega^{2}}{f}, \label{iz21}%
\end{equation}

which can be rewritten as a bi-quadratic equation as follows
\begin{equation}
a(\partial_{r}R)^{4}+b(\partial_{r}R)^{2}+c=0, \label{iz22}%
\end{equation}

where
\begin{equation}
a=-2\beta\frac{f^{2}}{(1+L)^{2}},
\end{equation}

\begin{equation}
b=\frac{f}{1+L}\left[  1-4\beta\left(  m^{2}+e\right)  \right]  ,
\end{equation}

\begin{equation}
c=e-2\beta e^{2}-4\beta em^{2}+m^{2}-\frac{\omega^{2}}{f}-2\beta m^{4}.
\end{equation}

Eq. (\ref{iz22}) has four roots if $b^{2}-4ac>0$. We deduced from our
analytical computations that only two roots ($R_{\pm}$) have physical meaning
at the event horizon of the SBHBGM. These roots are
\begin{equation}
R_{\pm}=\pm%
{\displaystyle\int}
dr\sqrt{\left(  1+L\right)  \frac{\omega^{2}-m^{2}f+2\beta m^{4}f}{f^{2}}%
}\left(  1+2\beta m^{2}\right)  =i\pi\omega M\sqrt{1+L}\left(  1+2\beta
m^{2}\right)  . \label{iz24}%
\end{equation}

It is worth noting that a $+/-$ sign represents an outgoing/ingoing wave. On
the other hand, the integrand of the integral (\ref{iz24}) has a pole at
$r=r_{h}$. Evaluating the integral by using the Cauchy's integral formula, we
obtain the imaginary part of the action as
\begin{equation}
\operatorname{Im}R_{\pm}\equiv\operatorname{Im}I_{\pm}=\pm\pi\omega
M\sqrt{1+L}\left(  1+2\beta m^{2}\right)  . \label{iz25}%
\end{equation}

Thus, the tunneling rate of the scalar particles becomes%
\begin{align}
\Gamma &  =\frac{P(emission)}{P(absorbtion)}=\frac{\exp(-2\operatorname{Im}%
I_{+})}{\exp(-2\operatorname{Im}I_{-})}=\exp(-4\operatorname{Im}%
I_{+})\nonumber\\
&  =\exp\left[  -4\omega M\pi\sqrt{1+L}\left(  1+2\beta m^{2}\right)  \right]
. \label{iz26n}%
\end{align}

Recalling the expression of the Boltzmann factor
\begin{equation}
\Gamma=\exp(-\frac{\omega}{T}), \label{iz27}%
\end{equation}

one can read the modified Hawking temperature ($\widetilde{T}_{H}$) as follows%
\begin{equation}
\widetilde{T}_{H}=\frac{1}{8\pi M\sqrt{1+L}\left[  1+2\beta m^{2}\right]
}=\frac{T_{H}}{(1+2\beta m^{2})}. \label{iz28}%
\end{equation}

As can be seen above, after terminating the GUP parameter i.e., $\beta=0$, one
can recover the standard Hawking temperature (\ref{iz13}).

\section{GUP-ASSISTED HR OF SBHBGM: FERMIONS' TUNNELING}

In this section, we aim to derive the modified Hawking temperature in the case
of radiating fermions. To this end, we consider the Dirac equation, which is
given by \cite{37}
\begin{equation}
\left\{  i\,\hbar\,\gamma^{0}\partial_{0}+\left[  m+i\,\hbar\,\gamma^{\mu
}\left(  \Omega_{\mu}+\hbar\,\beta\,\partial_{\mu}\right)  \right]  \left(
1-\beta\,m^{2}+\beta\,\hbar^{2}\,g_{jk}\,\partial^{j}\,\partial^{k}\right)
\right\}  \psi=0, \label{iz29n}%
\end{equation}

where $\psi$ denotes the test spinor field. The $\gamma^{\mu}$ matrices for
the metric ({\ref{iz9}}) are given by%

\begin{equation}%
\begin{tabular}
[c]{|c|c|}\hline
$\gamma^{t}=\frac{1}{\sqrt{f(r)}}\left(
\begin{array}
[c]{cc}%
i & 0\\
0 & -i
\end{array}
\right)  $ & $\gamma^{r}=\sqrt{\frac{f(r)}{1+L}}\left(
\begin{array}
[c]{cc}%
0 & \sigma^{3}\\
\sigma^{3} & 0
\end{array}
\right)  $\\\hline
$\gamma^{\theta}=\frac{1}{r}\left(
\begin{array}
[c]{cc}%
0 & \sigma^{1}\\
\sigma^{1} & 0
\end{array}
\right)  $ & $\gamma^{\phi}=\frac{1}{r\sin\theta}\left(
\begin{array}
[c]{cc}%
0 & \sigma^{2}\\
\sigma^{2} & 0
\end{array}
\right)  $\\\hline
\end{tabular}
\ \ \ \ , \label{iz30n}%
\end{equation}
in which $\sigma^{i}$'s represent the well-known Pauli matrices \cite{39}. One
can easily ignore the terms having $\beta^{2}$ since $\beta$ is the effect of
quantum gravity and it is a relatively very small quantity. For spin-up
particles, the wave function can be expressed as \cite{37}%

\begin{equation}
\Psi=\left(
\begin{array}
[c]{c}%
0\\
X\\
0\\
Y
\end{array}
\right)  \exp\left(  \frac{i}{\hbar}I\right)  , \label{iz31}%
\end{equation}

where $X$, $Y,$ and $I$ are functions of coordinates $\left(  t,r,\theta
,\phi\right)  .$ $I$ is the action of the emitted fermion. It is worth noting
that here we only consider the spin-up case since it is physically same with
the spin-down case; the only difference is the sign. Substitution of the wave
function in the generalized Dirac equation (\ref{iz29n}) results in the
following coupled equations%

\begin{multline}
-iX\frac{1}{\sqrt{f}}\partial_{t}I-Y(1-\beta m^{2})\sqrt{\frac{f}{1+L}%
}\partial_{r}I-Xm\beta\left[  \frac{f}{1+L}\left(  \partial_{r}I\right)
^{2}+g^{\theta\theta}\left(  \partial_{\theta}I\right)  ^{2}+g^{\varphi
\varphi}\left(  \partial_{\varphi}I\right)  ^{2}\right]  +\\
Y\beta\sqrt{\frac{f}{1+L}}\partial_{r}I\left[  \frac{f}{1+L}\left(
\partial_{r}I\right)  ^{2}+g^{\theta\theta}\left(  \partial_{\theta}I\right)
^{2}+g^{\phi\phi}\left(  \partial_{\varphi}I\right)  ^{2}\right]  +Xm\left(
1-\beta m^{2}\right)  =0, \label{034}%
\end{multline}

and%

\begin{multline}
iY\frac{1}{\sqrt{f}}\partial_{t}I-X\left(  1-\beta m^{2}\right)  \sqrt
{\frac{f}{1+L}}\partial_{r}I-Ym\beta\left[  \frac{f}{1+L}\left(  \partial
_{r}I\right)  ^{2}+g^{\theta\theta}\left(  \partial_{\theta}I\right)
^{2}+g^{\varphi\varphi}\left(  \partial_{\varphi}I\right)  ^{2}\right]  +\\
X\beta\sqrt{\frac{f}{1+L}}\partial_{r}I\left[  \frac{f}{1+L}\left(
\partial_{r}I\right)  ^{2}+g^{\theta\theta}\left(  \partial_{\theta}I\right)
^{2}+g^{\varphi\varphi}\left(  \partial_{\varphi}I\right)  ^{2}\right]
+Ym\left(  1-\beta m^{2}\right)  =0. \label{035}%
\end{multline}

Then, one can get the following decoupled equations%

\begin{multline}
X\left\{  -\left(  1-\beta m^{2}\right)  \sqrt{g^{\theta\theta}}%
\partial_{\theta}I+\beta\sqrt{g^{\theta\theta}}\partial_{\theta}I\left[
\frac{f}{1+L}\left(  \partial_{r}I\right)  ^{2}+g^{\theta\theta}\left(
\partial_{\theta}I\right)  ^{2}+g^{\varphi\varphi}\left(  \partial_{\varphi
}I\right)  ^{2}\right]  \right. \\
\left.  -i\left(  1-\beta m^{2}\right)  \sqrt{g^{\varphi\varphi}}%
\partial_{\varphi}I+i\beta\sqrt{g^{\varphi\varphi}}\partial_{\varphi}I\left[
\frac{f}{1+L}\left(  \partial_{r}I\right)  ^{2}+g^{\theta\theta}\left(
\partial_{\theta}I\right)  ^{2}+g^{\varphi\varphi}\left(  \partial_{\varphi
}I\right)  ^{2}\right]  \right\}  =0, \label{036}%
\end{multline}

and%

\begin{multline}
Y\left\{  -\left(  1-\beta m^{2}\right)  \sqrt{g^{\theta\theta}}%
\partial_{\theta}I+\beta\sqrt{g^{\theta\theta}}\partial_{\theta}I\left[
\frac{f}{1+L}\left(  \partial_{r}I\right)  ^{2}+g^{\theta\theta}%
(\partial_{\theta}I)^{2}+g^{\phi\phi}(\partial_{\phi}I)^{2}\right]  \right. \\
\left.  -i\left(  1-\beta m^{2}\right)  \sqrt{g^{\phi\phi}}\partial_{\phi
}I+i\beta\sqrt{g^{\phi\phi}}\partial_{\phi}I\left[  \frac{f}{1+L}\left(
\partial_{r}I\right)  ^{2}+g^{\theta\theta}\left(  \partial_{\theta}I\right)
^{2}+g^{\varphi\varphi}\left(  \partial_{\phi}I\right)  ^{2}\right]  \right\}
=0. \label{037}%
\end{multline}

By using the fact that SBHBGM spacetime has a timelike Killing vector
$\frac{\partial}{\partial t}$, one can obtain the radial action by performing
the separation of variables technique:
\begin{equation}
I=-\omega t+W(r)+\Theta(\theta,\phi), \label{iz36n}%
\end{equation}
where $\omega$ is the fermion energy. Substituting Eq. (\ref{iz36n}) into Eqs.
(\ref{036}) and (\ref{037}), we find out the identical equations for $X$ and
$Y$ equations. Thus, we have
\begin{multline}
\beta\left[  \frac{f}{1+L}\left(  \partial_{r}W\right)  ^{2}+g^{\theta\theta
}(\partial_{\theta}\Theta)^{2}+g^{\phi\phi}(\partial_{\phi}\Theta)^{2}\left(
\sqrt{g^{\theta\theta}}\partial_{\theta}\Theta+i\sqrt{g^{\varphi\varphi}%
}\partial_{\phi}\Theta\right)  \right]  +\\
\left(  1-\beta m^{2}\right)  \left(  \sqrt{g^{\theta\theta}}\partial_{\theta
}\Theta+i\sqrt{g^{\varphi\varphi}}\partial_{\varphi}\Theta\right)  =0,
\end{multline}

or%

\begin{equation}
\left(  \sqrt{g^{\theta\theta}}\partial_{\theta}\Theta+i\sqrt{g^{\varphi
\varphi}}\partial_{\varphi}\Theta\right)  \left[  \beta\left(  \frac{f}%
{1+L}\left(  \partial_{r}W\right)  ^{2}+g^{\theta\theta}(\partial_{\theta
}\Theta)^{2}+g^{\phi\phi}(\partial_{\phi}\Theta)^{2}+m^{2}\right)  -1\right]
=0. \label{iz38n}%
\end{equation}

It is obvious that the expression inside the square brackets can not vanish;
thus, one should have%

\begin{equation}
\left(  \sqrt{g^{\theta\theta}}\partial_{\theta}\Theta+i\sqrt{g^{\varphi
\varphi}}\partial_{\varphi}\Theta\right)  =0, \label{iz39n}%
\end{equation}

and the solution of $\Theta$, therefore, does not contribute to the tunneling
rate. The above result helps us to simplify Eqs. (\ref{034}) and (\ref{035})
[with ansatz (\ref{iz36n})] as follows%

\begin{multline}
X\left\{  \frac{i\omega}{\sqrt{f}}-m\beta\left[  \frac{f}{1+L}\left(
\partial_{r}W\right)  ^{2}\right]  +m\left(  1-\beta m^{2}\right)  \right\}
+\\
Y\left\{  -\left(  1-\beta m^{2}\right)  \sqrt{\frac{f}{1+L}}\partial
_{r}W+\beta\sqrt{\frac{f}{1+L}}\partial_{r}W\left[  \frac{f}{1+L}\left(
\partial_{r}W\right)  ^{2}\right]  \right\}  =0,
\end{multline}

and%

\begin{multline}
Y\left\{  \frac{i\omega}{\sqrt{f}}-m\beta\left[  \frac{f}{1+L}\left(
\partial_{r}W\right)  ^{2}\right]  +m\left(  1-\beta m^{2}\right)  \right\}
+\label{iz41n}\\
X\left\{  -\left(  1-\beta m^{2}\right)  \sqrt{\frac{f}{1+L}}\partial
_{r}W+\beta\sqrt{\frac{f}{1+L}}\partial_{r}W\left[  \frac{f}{1+L}\left(
\partial_{r}W\right)  ^{2}\right]  \right\}  =0.
\end{multline}

In the simple way, one can set%

\begin{equation}
XA+YB=0, \label{iz42}%
\end{equation}

\begin{equation}
YA+XB=0, \label{iz43}%
\end{equation}

where%

\begin{equation}
A=\frac{i\omega}{\sqrt{f}}-m\left[  \frac{f\beta}{1+L}\left(  \partial
_{r}W\right)  ^{2}+1-\beta m^{2}\right]  , \label{iz44}%
\end{equation}

and%

\begin{equation}
B=-\left(  1-\beta m^{2}\right)  \sqrt{\frac{f}{1+L}}\partial_{r}W+\beta
\sqrt{\frac{f}{1+L}}\partial_{r}W\left[  \frac{f}{1+L}\left(  \partial
_{r}W\right)  ^{2}\right]  . \label{iz45}%
\end{equation}

After making some manipulations, we see that $A^{2}-B^{2}=0$%

\begin{multline}
\left(  \frac{i\omega}{\sqrt{f}}-m\beta\left[  \frac{f}{1+L}\left(
\partial_{r}W\right)  ^{2}\right]  +m(1-\beta m^{2})\right)  ^{2}%
-\label{iz46}\\
\left(  -\left(  1-\beta m^{2}\right)  \sqrt{\frac{f}{1+L}}\partial_{r}%
W+\beta\sqrt{\frac{f}{1+L}}\partial_{r}W\left[  \frac{f}{1+L}\partial_{r}%
^{2}W\right]  \right)  ^{2}=0,
\end{multline}

which yields%

\begin{equation}
L_{6}\left(  \partial_{r}W\right)  ^{6}+L_{4}\left(  \partial_{r}W\right)
^{4}+L_{2}\left(  \partial_{r}W\right)  ^{2}+L_{0}=0, \label{iz47}%
\end{equation}

where%

\begin{equation}
L_{6}=\beta^{2}f(\frac{f}{1+L})^{3}, \label{iz48}%
\end{equation}

\begin{equation}
L_{4}=\beta(\frac{f}{1+L})^{2}f\left(  m^{2}\beta-2\right)  , \label{iz49}%
\end{equation}

\begin{equation}
L_{2}=\frac{f^{2}}{1+L}\left(  \frac{2i\omega m}{\sqrt{f}}+\left(  1-\beta
m^{2}\right)  \left(  1+2m^{2}\beta\right)  \right)  , \label{iz50}%
\end{equation}

\begin{equation}
L_{0}=\omega^{2}-m^{2}f\left(  1-\beta m^{2}\right)  ^{2}-2i\omega m\sqrt
{f}\left(  1-\beta m^{2}\right)  , \label{iz51}%
\end{equation}

ignoring $O(\beta^{2})$ terms, Eq. (\ref{iz47}) reduces to%

\begin{equation}
L_{4}\left(  \partial_{r}W\right)  ^{4}+L_{2}\left(  \partial_{r}W\right)
^{2}+L_{0}=0. \label{iz52}%
\end{equation}

Therefore, we have%

\begin{align}
W_{\pm}  &  =\pm\int dr\frac{\sqrt{\left(  1+L\right)  \left(  \omega
^{2}+m^{2}f\right)  }}{f}\left[  1+\beta\left(  m^{2}+\frac{\omega^{2}}%
{f}\right)  \right] \nonumber\\
&  \cong\pm i\pi\omega r_{+}\left(  1+2\beta m^{2}\right) \label{iz53}\\
&  =\pm i2\pi M\omega\sqrt{1+L}\left(  1+2\beta m^{2}\right)  ,\nonumber
\end{align}

in another form%

\begin{equation}
\operatorname{Im}W_{\pm}=2\pi M\omega\sqrt{1+L}\left(  1+2\beta m^{2}\right)
. \label{iz54}%
\end{equation}

Recalling Eq. (\ref{iz26n}), we find the tunneling rate of fermions as
follows
\begin{equation}
\Gamma\simeq\exp\left(  -4\,\mathrm{Im}W_{+}\right)  =\exp\left(  8\pi
M\omega\sqrt{1+L}\left(  1+2\beta m^{2}\right)  \right)  . \label{iz55}%
\end{equation}

Thus, with the help of the Boltzmann factor (\ref{iz27}), we get the
GUP-consolidated temperature of the SBHBGM via the emission of the fermions:%

\begin{equation}
T=\frac{1}{8\pi M\sqrt{1+L}\left(  1+2\beta m^{2}\right)  }=\frac{T_{0}%
}{\left(  1+2\beta m^{2}\right)  }, \label{iz56}%
\end{equation}

in which $T_{0}$ represents the original Hawking temperature (\ref{iz13})%

\begin{equation}
T_{0}=\frac{1}{8\pi M\sqrt{1+L}}. \label{iz57}%
\end{equation}

The above result \eqref{iz56} shows that GUP corrected temperature deviates
from the standard Hawking temperature.

\section{Greybody Factors of SBHBGM}

In this section, we shall first derive the effective potentials of the scalar
and fermion perturbations in the geometry of the SBHBGM. Then, the obtained
effective potentials will be used for computing the greybody factors of the
SBHBGM. The results will be depicted with some plots and discussed.

\subsection{Scalar Perturbations of SBHBGM}

The massless Klein- Gordon equation is given by%

\begin{equation}
\square\Psi=0, \label{iz58}%
\end{equation}

where the D'Alembert operator is denoted by the box symbol and $\square
=\frac{1}{\sqrt{-g}}\partial_{\mu}(\sqrt{-g}g^{\mu\nu}\partial_{\nu})$. For
the SBHBGM (\ref{iz9}), we have%

\begin{equation}
\sqrt{-g}=r^{2}\sin\theta\sqrt{1+L}, \label{iz59}%
\end{equation}

and therefore Eq. (\ref{iz58}) reads%

\begin{multline}
\square\Psi=\frac{1}{f}\partial_{t}^{2}-\frac{1}{r^{2}(1+L)}\left(
2rf\partial_{r}\Psi+r^{2}\partial_{r}f\partial r\Psi+r^{2}f\partial_{r}%
^{2}\Psi\right)  +\nonumber\\
\frac{1}{r^{2}\sin\theta}(-\cos\theta\partial_{\theta}\Psi-\sin\theta
\partial_{\theta}^{2}\Psi)-\frac{1}{r^{2}\sin^{2}\theta}\partial_{\phi}%
^{2}\Psi. \label{iz60n}%
\end{multline}

We invoke the following ansatz for the scalar field $\Psi$\ in the above equation:%

\begin{equation}
\Psi=p(r)A(\theta)e^{-i\omega t}e^{im\phi}, \label{iz61}%
\end{equation}

so that we have%

\begin{multline}
\square\Psi=-\frac{\omega^{2}}{f}-\frac{1}{p(1+L)}\left[  \frac{2f}%
{r}p^{\prime}+f^{\prime}p^{\prime}+fp^{\prime\prime}\right]  -\nonumber\\
\frac{1}{r^{2}A\sin\theta}\left[  \cos\theta A^{\prime}+\sin\theta
A^{\prime\prime}-\frac{m^{2}}{\sin\theta}A\right]  =0. \label{iz62n}%
\end{multline}

When one changes the independent variable $\theta$ to $\cos^{-1}z$, the
angular equation is found to be%

\begin{equation}
\left(  1-z^{2}\right)  A^{\prime\prime}+2zA^{\prime}-\left[  m^{2}%
+\frac{\lambda_{s}}{1+L}\left(  1-z^{2}\right)  \right]  A=0, \label{iz63}%
\end{equation}

where $\lambda_{s}$ denotes the eigenvalue. The above equation is nothing but
the Legendre differential equation when one sets%

\begin{equation}
\lambda_{s}=-l(l+1)(1+L). \label{iz65}%
\end{equation}

The radial equation then becomes%

\begin{equation}
p^{\prime\prime}+p^{\prime}\left(  \frac{f^{\prime}}{f}+\frac{2}{r}\right)
+\left[  \frac{1+L}{f^{2}}\omega^{2}+\frac{\lambda_{s}}{r^{2}f}\right]  p=0.
\label{iz66}%
\end{equation}

Introducing a new variable $p=\frac{u}{r},$ we get a Schr\"{o}dinger-like wave equation%

\begin{equation}
\frac{du^{2}}{dr_{\ast}^{2}}+\left(  \omega^{2}-V_{eff}\right)  u=0,
\label{iz67}%
\end{equation}

where $r_{\ast}$ is the tortoise coordinate defined by%

\begin{equation}
r_{\ast}=\sqrt{1+L}%
{\displaystyle\int}
\frac{dr}{f}. \label{iz68}%
\end{equation}

The effective potential felt by the scalar field then becomes%

\begin{equation}
V_{eff}=f\left[  \frac{f^{\prime}}{\left(  1+L\right)  r}+\frac{l(l+1)}{r^{2}%
}\right]  . \label{iz69}%
\end{equation}

\begin{figure}[h]
\centering
\includegraphics[scale=.5]{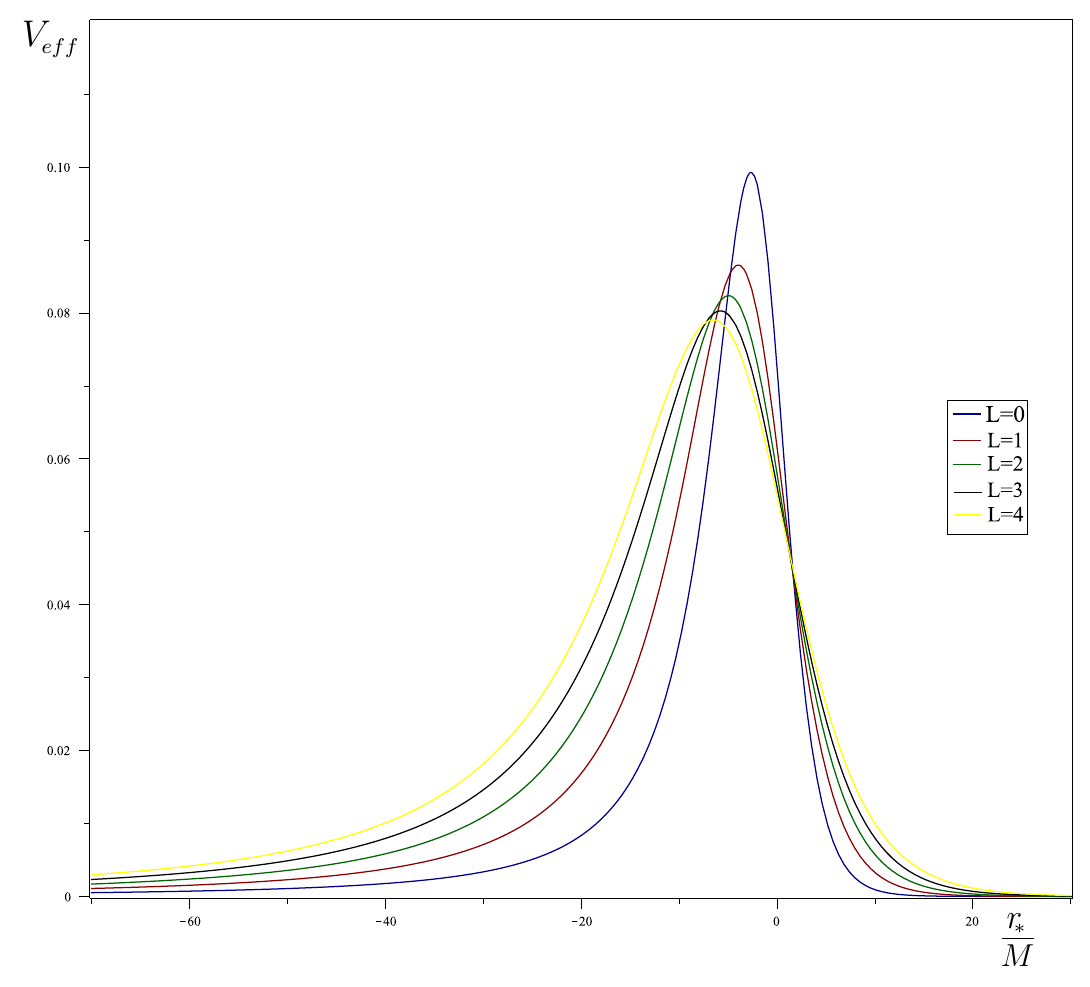}\caption{$V_{eff}$ versus $\frac{r_{\ast
}}{M}$ graph. The plots are governed by Eq. (\ref{iz69}).}%
\label{fig1}%
\end{figure}

It is obvious from Fig. (\ref{fig1}) that the effective potential vanishes
both at the event horizon of the SBHBG$T$ and at spatial infinity. This
behavior will help us to analytically derive the greybody factor of the scalar
field emission from the SBHBG$T$.

\subsection{Fermion Perturbations of SBHBGM}

In this subsection, we shall employ the Newman-Penrose formalism \cite{40} to
find the effective potential of the fermion fields propagating in the geometry
of the SBHBGM. Chandrasekar-Dirac equations (CDEs) are given by
\cite{Chandram}%
\begin{equation}
\left(  D+\varepsilon-\rho\right)  F_{1}+(\overline{\delta}+\pi-\alpha
)F_{2}=i\mu^{\ast}G_{1},\nonumber
\end{equation}

\begin{equation}
\left(  \delta+\beta-\tau\right)  F_{1}+\left(  \Delta+\mu-\gamma\right)
F_{2}=i\mu^{\ast}G_{2},\nonumber
\end{equation}

\begin{equation}
\left(  D+\overline{\varepsilon}-\overline{\rho}\right)  G_{2}-\left(
\delta+\overline{\pi}-\overline{\alpha}\right)  =i\mu^{\ast}F_{2},\nonumber
\end{equation}

\begin{equation}
\left(  \Delta+\overline{\mu}-\overline{\gamma}\right)  G_{1}-\left(
\overline{\delta}+\overline{\beta}-\overline{\tau}\right)  G_{2}=i\mu^{\ast
}F_{1}, \label{iz70}%
\end{equation}

where $F_{1},$ $F_{2},$ $G_{1},$ and $G_{2}$ represent the components of the
wave functions or the so-called Dirac spinors. $\varepsilon,$ $\rho$, $\pi,$
$\alpha,$ $\beta,$ $\tau,$ $\mu,$ and $\gamma$ are the spin coefficients, and
a bar over a quantity denotes complex conjugation. The non-zero spin
coefficients are found to be%

\begin{equation}
\varepsilon=\gamma=\frac{\sqrt{2}f^{\prime}}{8\sqrt{f}\sqrt{1+L}},\text{
\ \ \ \ \ \ }\mu=\rho=-\frac{\sqrt{2f}}{2r\sqrt{1+L}},\text{ \ \ \ \ \ }%
\beta=-\alpha=-\frac{\sqrt{2}\cot\theta}{4r}. \label{iz71}%
\end{equation}

To have separable solutions for the CDEs (\ref{iz70}), we introduce the
following ansatzes%

\begin{equation}
F_{1}=f_{1}(z)A_{1}(\theta)\exp\left[  i(\omega t+m\phi)\right]  ,\nonumber
\end{equation}

\begin{equation}
G_{1}=g_{1}(z)A_{2}(\theta)\exp\left[  (\omega t+m\phi)\right]  ,\nonumber
\end{equation}

\begin{equation}
F_{2}=f_{2}(z)A_{3}(\theta)\exp\left[  i(\omega t+m\phi)\right]  ,\nonumber
\end{equation}

\begin{equation}
G_{2}=g_{2}(z)A_{4}(\theta)\exp\left[  (\omega t+m\phi)\right]  , \label{iz72}%
\end{equation}

where $m$ denotes the azimuthal number and $\omega$ is the frequency of the
spinor fields. Since the directional derivatives \cite{Chandram} are defined
by $D=\ell^{a}\partial_{a},$ $\Delta=n^{a}\partial_{a},$ and $\delta
=m^{a}\partial_{a},$ we have%

\begin{equation}
D=\frac{1}{\sqrt{2f}}\partial_{t}+\sqrt{\frac{f}{2\left(  1+L\right)  }%
}\partial_{r},\nonumber
\end{equation}

\begin{equation}
\Delta=\frac{1}{\sqrt{2f}}\partial_{t}-\sqrt{\frac{f}{2(1+L)}}\partial
_{r},\nonumber
\end{equation}

\begin{equation}
\delta=\frac{1}{r\sqrt{2}}\partial_{\theta}+\frac{i}{r\sqrt{2}\sin\theta
}\partial_{\varphi},\nonumber
\end{equation}

\begin{equation}
\overline{\delta}=\frac{1}{r\sqrt{2}}\partial_{\theta}-\frac{i}{r\sqrt{2}%
\sin\theta}\partial_{\varphi}. \label{iz73}%
\end{equation}

After substituting Eqs. (\ref{iz71}-\ref{iz73}) into the CDEs (\ref{iz70}),
one can obtain the following set of equations:%

\begin{equation}
\left[  \frac{i\omega}{\sqrt{f}}+\frac{r\sqrt{f}}{\sqrt{1+L}}\partial
_{r}+\frac{rf^{\prime}}{4\sqrt{f\left(  1+L\right)  }}+\frac{\sqrt{f}}%
{\sqrt{1+L}}\right]  \frac{f_{1}}{f_{2}}+\frac{\overset{\backsim}{L}A_{3}%
}{A_{1}}-i\mu r\frac{g_{1}A_{2}}{f_{2}A_{1}}=0,\nonumber
\end{equation}

\begin{equation}
\left[  \frac{i\omega}{\sqrt{f}}-\frac{r\sqrt{f}}{\sqrt{1+L}}\partial
_{r}-\frac{rf^{\prime}}{4\sqrt{f\left(  1+L\right)  }}-\frac{\sqrt{f}}%
{\sqrt{1+L}}\right]  \frac{f_{2}}{f_{1}}+\frac{\overset{\backsim}{L^{\dagger}%
}A_{1}}{A_{3}}-i\mu r\frac{g_{2}A_{4}}{f_{1}A_{3}}=0,\nonumber
\end{equation}

\begin{equation}
\left[  \frac{i\omega}{\sqrt{f}}+\frac{r\sqrt{f}}{\sqrt{1+L}}\partial
_{r}+\frac{rf^{\prime}}{4\sqrt{f\left(  1+L\right)  }}+\frac{\sqrt{f}}%
{\sqrt{1+L}}\right]  \frac{g_{2}}{g_{1}}-\frac{\overset{\backsim}{L^{\dagger}%
}A_{2}}{A_{4}}-i\mu r\frac{f_{2}A_{3}}{g_{1}A_{4}}=0,\nonumber
\end{equation}

\begin{equation}
\left[  \frac{i\omega}{\sqrt{f}}-\frac{r\sqrt{f}}{\sqrt{1+L}}\partial
_{r}-\frac{rf^{\prime}}{4\sqrt{f\left(  1+L\right)  }}-\frac{\sqrt{f}}%
{\sqrt{1+L}}\right]  \frac{g_{1}}{g_{2}}-\frac{\overset{\backsim}{L}A_{4}%
}{A_{2}}-i\mu r\frac{f_{1}A_{1}}{g_{2}A_{2}}=0, \label{iz74}%
\end{equation}

where $\mu=\sqrt{2}\mu^{\ast}.$ $\overset{\backsim}{L}$ and $\overset{\backsim
}{L^{\dagger}}$ are the angular operators, which are known as the laddering operators:%

\begin{equation}
\overset{\backsim}{L}=\partial_{\theta}+\frac{m}{\sin\theta}+\frac{\cot\theta
}{2},\text{ \ \ \ \ \ \ \ \ \ }\overset{\backsim}{L^{\dagger}}=\partial
_{\theta}-\frac{m}{\sin\theta}+\frac{\cot\theta}{2}, \label{iz75}%
\end{equation}

which lead to the spin-weighted spheroidal harmonics \cite{SWSH1,SWSH2} with
the following eigenvalue \cite{myTeukolsky,LeePRD}:%

\begin{equation}
\lambda_{f}=-\left(  l+\frac{1}{2}\right)  . \label{iz75n2}%
\end{equation}

By considering $g_{1}=f_{2},$ $g_{2}=f_{1,}$ $A_{2}=A_{1,\text{ }}$and
$A_{4}=A_{3},$ then we reduce the CDEs (\ref{iz74}) to two coupled
differential equations:%

\begin{equation}
\frac{r\sqrt{f}}{\sqrt{1+L}}\left(  \frac{d}{dr}+\frac{i\omega\sqrt{1+L}}%
{f}+\frac{f^{\prime}}{4f}+\frac{1}{r}\right)  g_{2}=\left(  -\lambda_{f}+i\mu
r\right)  g_{1}, \label{iz76}%
\end{equation}

\begin{equation}
\frac{r\sqrt{f}}{\sqrt{1+L}}\left(  \frac{d}{dr}-\frac{i\omega\sqrt{1+L}}%
{f}+\frac{f^{\prime}}{4f}+\frac{1}{r}\right)  g_{1}=\left(  -\lambda_{f}-i\mu
r\right)  g_{2}. \label{iz77}%
\end{equation}

Moreover, if one sets%

\begin{equation}
g_{1}(r)=\frac{\Psi_{1}}{r}\text{, \ \ and \ \ }g_{2}(r)=\frac{\Psi_{2}}{r},
\label{iz78}%
\end{equation}

and substitute them into Eqs. (\ref{iz76}) and (\ref{iz77}), after some
manipulations, we get:%

\begin{equation}
\frac{r\sqrt{f}}{\sqrt{1+L}}\left(  \frac{d}{dr}+\frac{i\omega\sqrt{1+L}}%
{f}+\frac{f^{\prime}}{4f}\right)  \Psi_{2}=\left(  -\frac{\lambda_{f}}{r}%
+i\mu\right)  \Psi_{1}, \label{iz79}%
\end{equation}

\begin{equation}
\frac{r\sqrt{f}}{\sqrt{1+L}}\left(  \frac{d}{dr}-\frac{i\omega\sqrt{1+L}}%
{f}+\frac{f^{\prime}}{4f}\right)  \Psi_{1}=\left(  -\frac{\lambda_{f}}{r}%
-i\mu\right)  \Psi_{2}. \label{iz80}%
\end{equation}

By defining $\Psi_{1}=f^{-\frac{1}{4}}R_{1}(r)$ and $\Psi_{2}=f^{-\frac{1}{4}%
}R_{2}(r)$ and introducing the tortoise coordinate ($r_{\ast}$) as $\frac
{f}{\sqrt{1+L}}\frac{d}{dr}=\frac{d}{dr_{\ast}}$, we obtain%

\begin{equation}
\left(  \frac{d}{dr_{\ast}}+i\omega\right)  R_{2}(r)=\sqrt{f}\left(
-\frac{\lambda_{f}}{r}+i\mu\right)  R_{1}, \label{iz81}%
\end{equation}

\begin{equation}
\left(  \frac{d}{dr_{\ast}}-i\omega\right)  R_{1}(r)=\sqrt{f}\left(
-\frac{\lambda_{f}}{r}-i\mu\right)  R_{2}. \label{iz82}%
\end{equation}

One can combine the above equations by letting%

\begin{equation}
Z_{+}=R_{1}+R_{2}, \label{iz83}%
\end{equation}

\begin{equation}
Z_{-}=R_{2}-R_{1}. \label{iz84}%
\end{equation}

Thus, we end up with the following pair of one dimensional
Schr\"{o}dinger-like wave equations:%

\begin{equation}
\left(  \frac{d^{2}}{dr_{\ast}^{2}}+\omega^{2}\right)  Z_{\pm}=V_{\pm}Z_{\pm},
\label{iz85}%
\end{equation}

where the effective potentials for the Dirac field read%

\begin{align}
V_{\pm}  &  =f\left[  \left(  -\frac{\lambda_{f}}{r}\pm i\mu\right)  ^{2}%
\pm\lambda_{f}\frac{1}{\sqrt{1+L}}\frac{d}{dr}\left(  -\frac{\sqrt{f}}{r}%
\pm\frac{i\mu\sqrt{f}}{\lambda_{f}}\right)  \right]  ,\label{iz86}\\
&  =f\left[  \left(  \frac{2l+1}{2r}\pm i\mu\right)  ^{2}\mp\left(  l+\frac
{1}{2}\right)  \frac{1}{\sqrt{1+L}}\frac{d}{dr}\left(  -\frac{\sqrt{f}}{r}%
\mp\frac{2i\mu\sqrt{f}}{2l+1}\right)  \right]  . \label{iz86n2}%
\end{align}

\begin{figure}[ptb]
\centering
\begin{subfigure}{.5\textwidth}
\centering
\includegraphics[scale=.421]{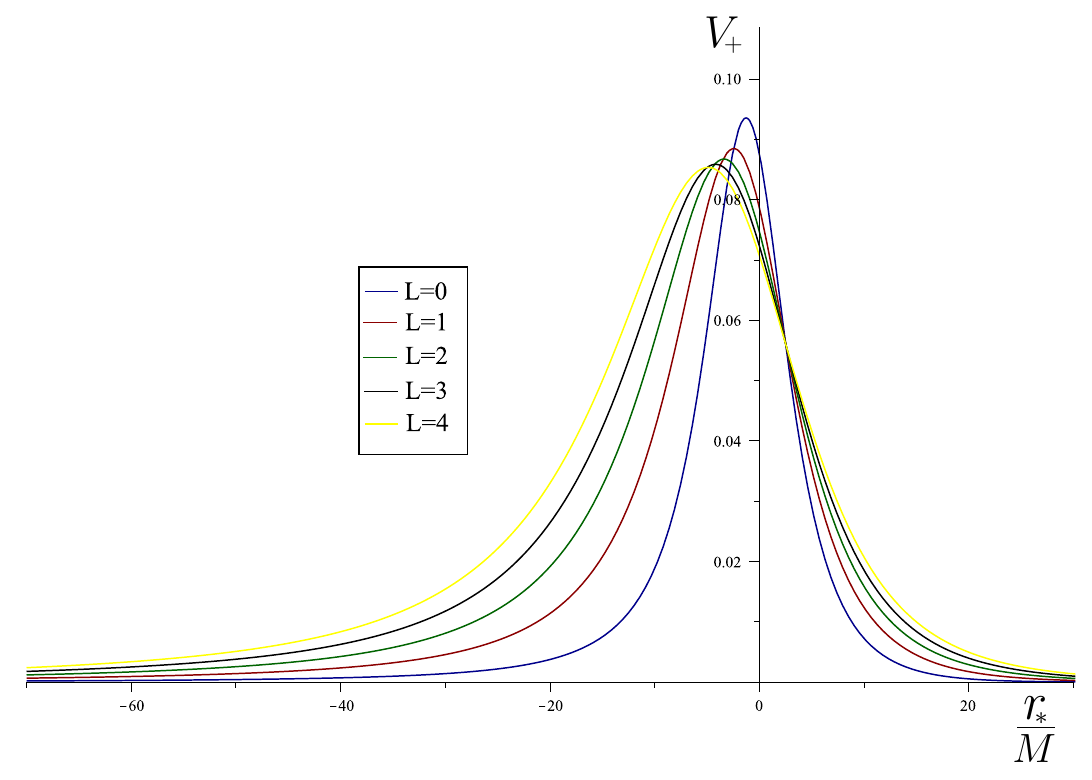}
\caption{$V_{+}$ versus $\frac{r_{\ast}}{M}$ graph.}%
\label{fig2a}%
\end{subfigure}\begin{subfigure}{.5\textwidth}
\centering
\includegraphics[scale=.4]{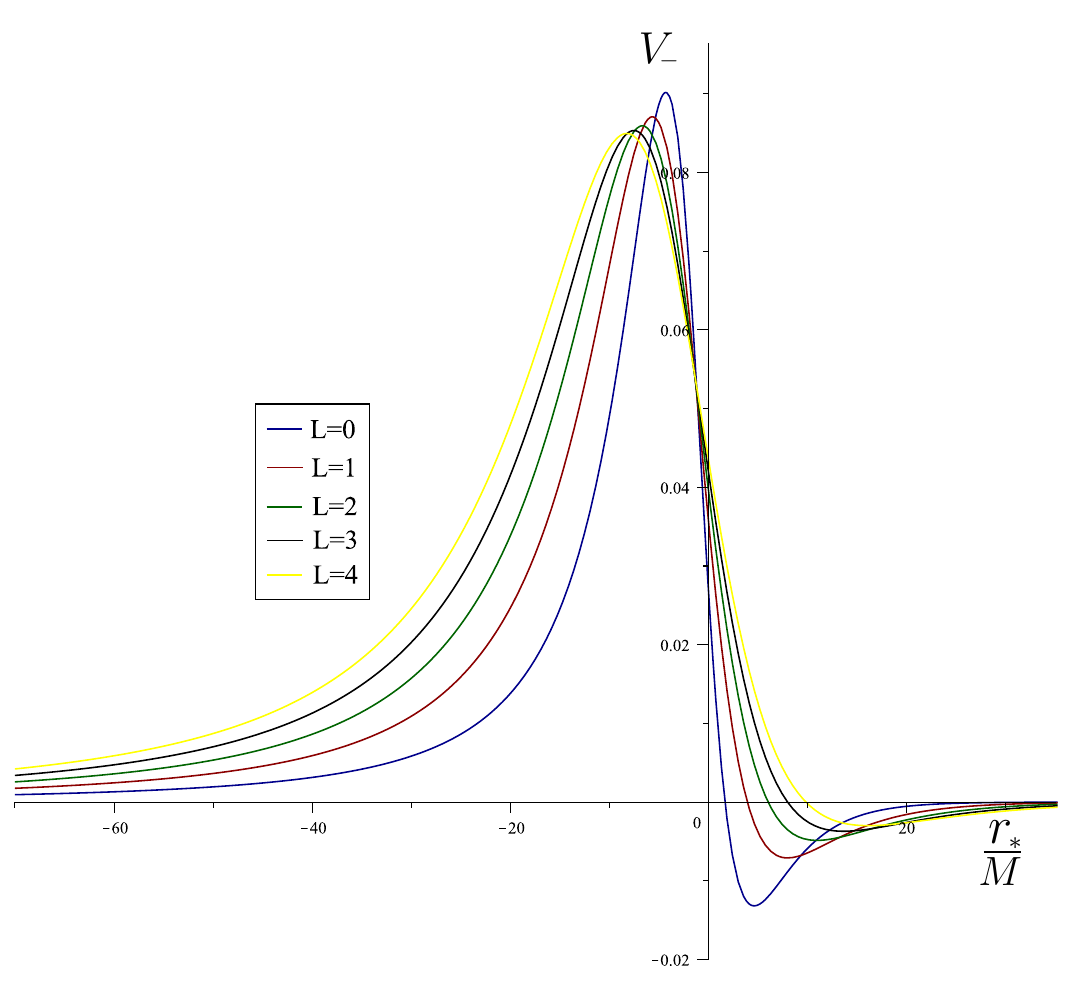}
	\caption{$V_{-}$ versus $\frac{r_{\ast}}{M}$ graph.}%
\label{fig2b}%
\end{subfigure}
\caption{Plots of $V_{\pm}$ versus $\frac{r_{\ast}}{M}$. The plots are
governed by Eq. (\ref{iz86n2}). }%
\label{fig2}%
\end{figure}

\subsection{Greybody Factor Computations}

In general relativity, the greybody factor is one of the most important
physical quantities related to the quantum nature of a BH. A high value of the
greybody factor indicates a high probability that HR can reach to spatial
infinity. Among the many methods (see for example \cite{WaldGBF}\ and
references therein) for obtaining the greybody factor, here we employ the
method of \cite{41}, which formulates the general semi-analytic bounds for
greybody factors:
\begin{equation}
\sigma_{\ell}(\omega)\geq\sec h^{2}\left(
{\displaystyle\int_{-\infty}^{+\infty}}
\wp dr_{\star}\right)  , \label{iz87}%
\end{equation}

where $\sigma_{\ell}(\omega)$ are the dimensionless greybody factors that
depend on the angular momentum quantum number $\ell$ and frequency $\omega$ of
the emitted particles, and
\begin{equation}
\wp=\frac{\sqrt{(h^{\prime})^{2}+(\omega^{2}-V_{eff}-h^{2})^{2}}}{2h},
\label{iz88}%
\end{equation}

in which prime denotes the derivation with respect to $r$. We have two
conditions for the certain positive function $h$: 1) $h(r_{\star})>0$ and 2)
$h(-\infty)=h(\infty)=\omega$ \cite{41}. Without loss of generality, we simply
set $h=\omega,$ which reduces the integration of Eq. (\ref{iz87}) to%

\begin{equation}%
{\displaystyle\int_{-\infty}^{+\infty}}
\wp dr_{\star}=\frac{\sqrt{1+L}}{2\omega}%
{\displaystyle\int_{r_{h}}^{+\infty}}
\frac{V_{eff}}{f(r)}dr. \label{iz89}%
\end{equation}

For a massless scalar field $\phi$, considering the effective potential given
in Eq. (\ref{iz69}), Eq. (\ref{iz87}) becomes
\begin{equation}
\sigma_{\ell}^{s}(\omega)\geq\sec h^{2}\left(  \frac{\sqrt{1+L}}{2\omega}%
\int_{r_{h}}^{+\infty}\left[  \frac{f^{\prime}(r)}{\left(  1+L\right)
r}+\frac{l(l+1)}{r^{2}}\right]  dr\right)  . \label{iz90}%
\end{equation}

Taking cognizance of the integral part of Eq. (\ref{iz90}):
\begin{align}
&  \frac{1}{2\omega}\int_{-\infty}^{+\infty}\left[  \frac{l(l+1)}{r^{2}%
}f(r)dr_{\star}+\frac{f\prime(r)}{r\left(  1+L\right)  }f(r)dr_{\star}\right]
,\nonumber\\
&  =\frac{\sqrt{1+L}}{2\omega}\left[  l(l+1)%
{\displaystyle\int_{r_{h}}^{+\infty}}
\frac{dr}{r^{2}}+\int_{r_{h}}^{\infty}\frac{dr}{r\left(  1+L\right)  }\left(
\frac{2M}{r^{2}}\right)  \right]  ,\nonumber\\
&  =\frac{\sqrt{1+L}}{2\omega r_{h}}\left[  l(l+1)+\frac{1}{2\left(
1+L\right)  }\right]  ,\text{ } \label{iz91}%
\end{align}

the greybody factor of the SBHBGM due to scalar field radiation yields%
\begin{equation}
\sigma_{\ell}^{s}(\omega)\geq\sec h^{2}\left\{  \frac{\sqrt{1+L}}{2\omega
r_{h}}\left[  l(l+1)+\frac{1}{2\left(  1+L\right)  }\right]  \right\}  .
\label{iz92n}%
\end{equation}

\begin{figure}[h]
\centering
\includegraphics[scale=.45]{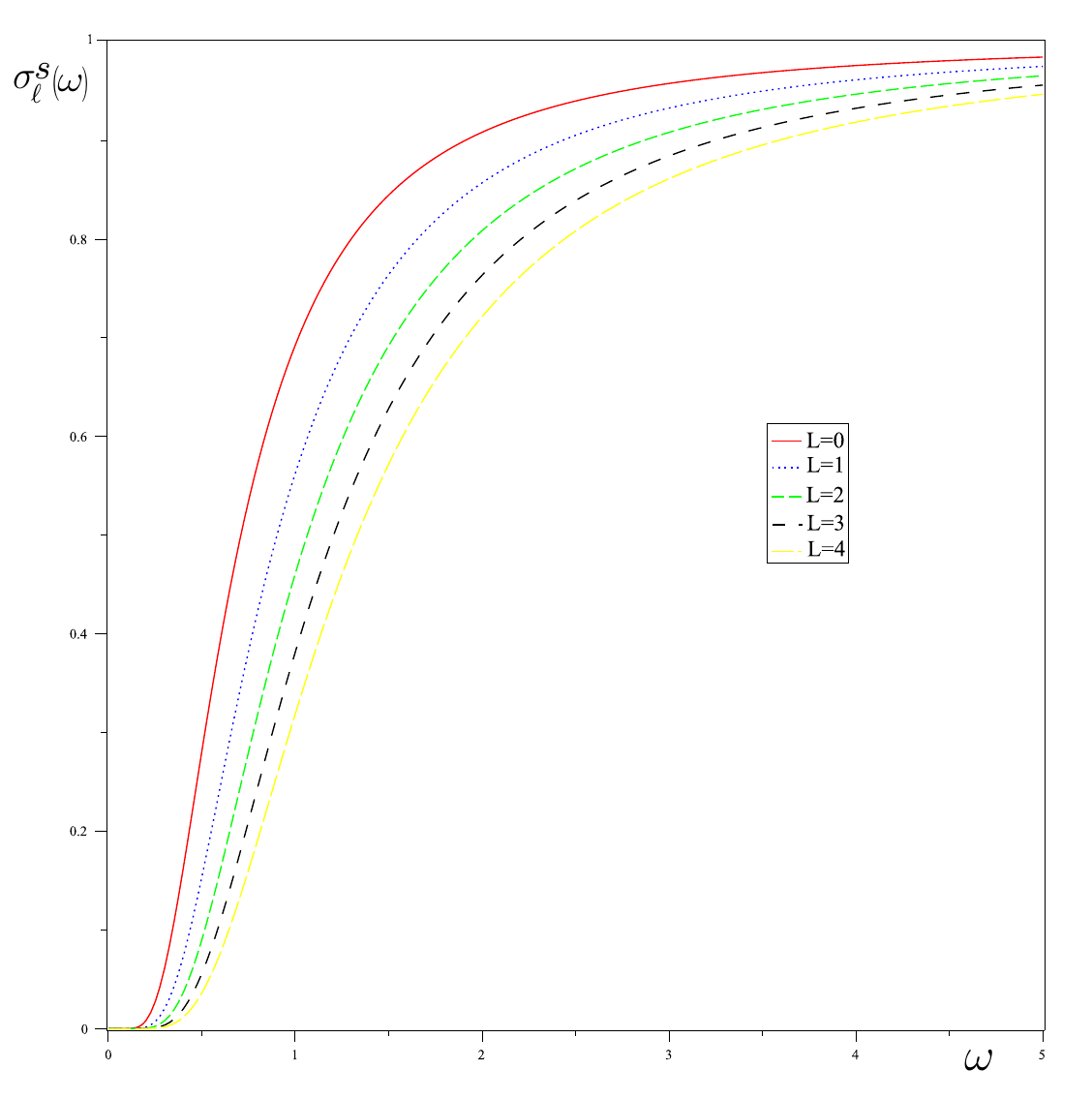}\caption{$\sigma_{\ell}^{s}(\omega)$
versus $\omega$ graph. The plots are governed by Eq. (\ref{iz92n}) with
$M=1$.}%
\label{fig3}%
\end{figure}

When one considers the effective potential (\ref{iz86}) of the massless Dirac fields:%

\begin{equation}
\left.  V_{\pm}\right\vert _{\mu=0}=f\left[  \frac{\lambda_{f}^{2}}{r^{2}}%
\pm\lambda_{f}\frac{1}{\sqrt{1+L}}\frac{d}{dr}\left(  -\frac{\sqrt{f}}%
{r}\right)  \right]  , \label{iz93}%
\end{equation}

the integral seen in Eq. (\ref{iz87}) can be easily computed. Thus, we find
the greybody factor expression of the SBHBGM arising from the fermion radiation:%

\begin{equation}
\sigma_{\ell}^{f}(\omega)\geq\sec h^{2}\left(  \frac{1}{2\omega}\int_{-\infty
}^{+\infty}f\left[  \frac{\lambda_{f}^{2}}{r^{2}}\pm\lambda_{f}\frac{1}%
{\sqrt{1+L}}\frac{d}{dr}\left(  -\frac{\sqrt{f}}{r}\right)  \right]
dr_{\star}\right)  . \label{iz94}%
\end{equation}

From now on, without loss of generality, we consider only $V_{+}$. After some
manipulation, one can get%

\begin{equation}
\sigma_{\ell}^{f}(\omega)\geq\sec h^{2}\left[  \frac{\sqrt{1+L}}{2\omega
}\left(  \lambda_{f}^{2}\int_{r_{h}}^{\infty}\left(  \frac{1}{r^{2}}\right)
dr\pm\frac{\lambda_{f}}{\sqrt{1+L}}\int_{r_{h}}^{\infty}\left(  1+\frac{M}%
{r}\right)  \left(  \frac{1}{r^{2}}-\frac{3M}{r^{3}}\right)  dr\right)
\right]  , \label{iz95}%
\end{equation}

which recasts in%

\begin{equation}
\sigma_{\ell}^{f}(\omega)\geq\sec h^{2}\left[  \frac{\sqrt{1+L}}{2\omega
}\left(  \lambda_{f}^{2}\left(  \frac{1}{r_{h}}\right)  \pm\frac{\lambda_{f}%
}{\sqrt{1+L}}\left[  \frac{1}{r_{h}}-\frac{M}{r_{h}^{2}}-\frac{M^{2}}%
{r_{h}^{3}}\right]  \right)  \right]  , \label{iz96}%
\end{equation}

or, in more compact form:%

\begin{equation}
\sigma_{\ell}^{f}(\omega)\geq\sec h^{2}\left[  \frac{\left(  l+\frac{1}%
{2}\right)  \sqrt{1+L}}{4M\omega}\left(  l+\frac{1}{2}\pm\frac{1}{4\sqrt{1+L}%
}\right)  \right]  . \label{iz97}%
\end{equation}

\begin{figure}[ptb]
\centering
\begin{subfigure}{.5\textwidth}
\centering
\includegraphics[scale=.4]{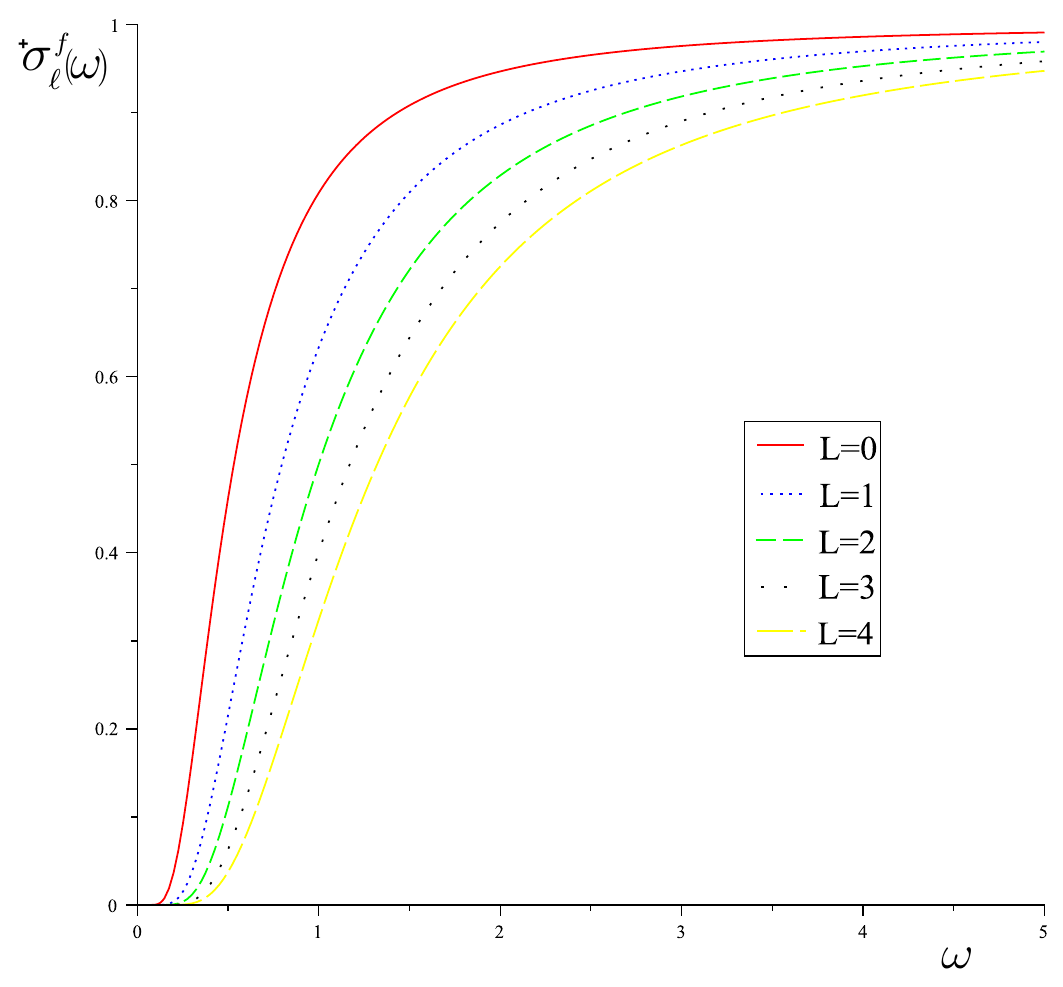}
\caption{$\sigma_{\ell}^{f}(\omega)$ versus $\omega$ graph for $V_{+}$.}%
\label{fig4a}%
\end{subfigure}\begin{subfigure}{.5\textwidth}
\centering
\includegraphics[scale=.4]{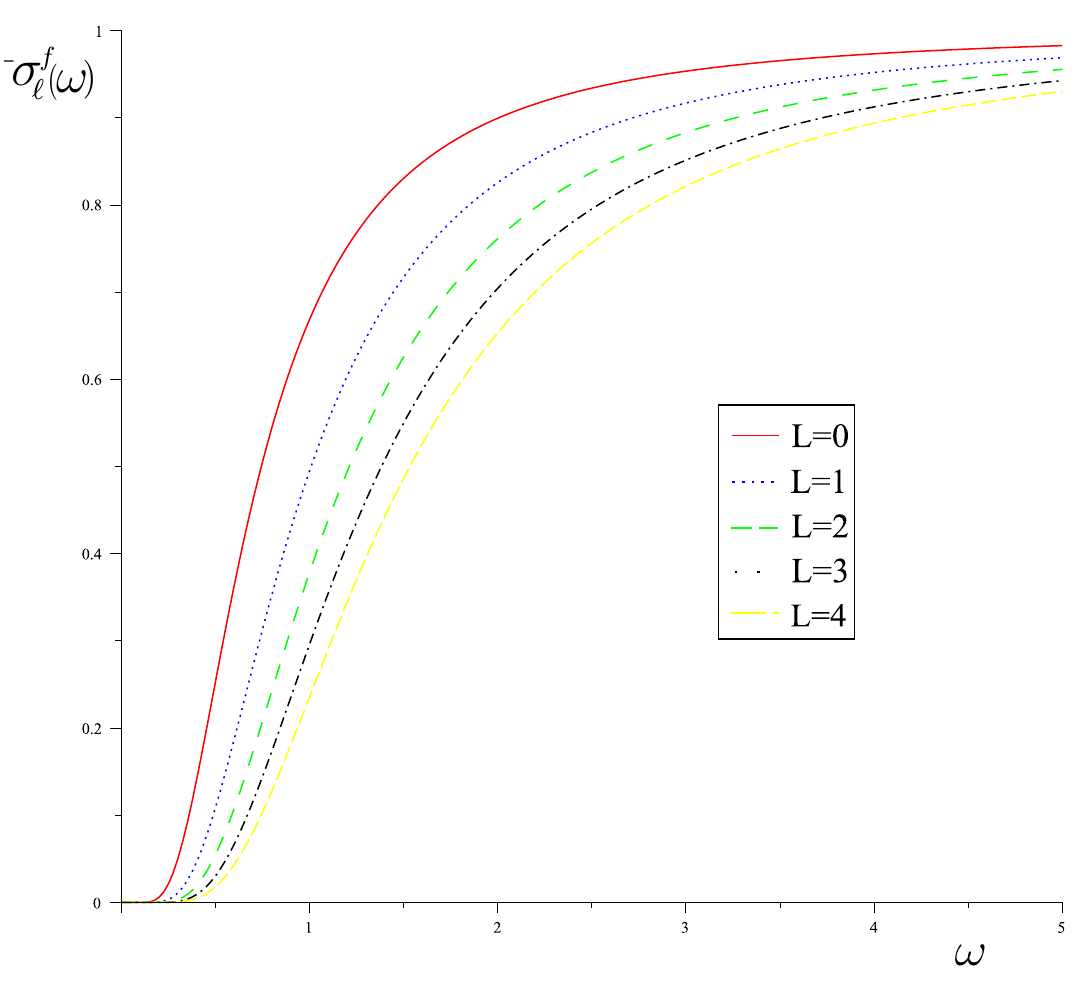}
	\caption{$\sigma_{\ell}^{f}(\omega)$ versus $\omega$ graph for $V_{-}$.}%
\label{fig4b}%
\end{subfigure}
\caption{Plots of $\sigma_{\ell}^{f}(\omega)$ versus $\omega$. The plots are
governed by Eq. (\ref{iz97}). }%
\label{fig4}%
\end{figure}

We depict the greybody factors of the SBHBGM arising from the scalar
(\ref{iz92n}) and fermion (\ref{iz97}) fields in Figs. (\ref{fig3}) and
(\ref{fig4}), respectively. As is well-known, the greybody factor of the HR
must be $<1$ since a BH does not perform a complete black body radiation with
a 100\% absorption coefficient. Our findings, as shown in Figs. (\ref{fig3})
and (\ref{fig4}), are in good agreement with the latter remark. Also, it can
be seen from these figures that the peak values of the greybody factors
decreases with increasing LSB parameter $L$. In summary, LSB has a greybody
factor-reducing effect.

\section{CONCLUSION}

In this paper, we studied the quantum thermodynamics \cite{QTerm} of the
SBHBGM. During this analysis, we had mainly two aims: 1) to obtain the
modified Hawking temperature of the SBHBGM, within the framework of GUP,
arising from the emission of bosons and fermions; 2) to compute the greybody
factors of the scalar and fermion fields from the SBHBGM. To this end, we
first derived the effective potentials of the Klein-Gordon and Dirac
equations. Next, we used the obtained effective potentials in the greybody
expression (\ref{iz87}). Then, we illustrated the obtained greybody factors in
Figs. (3) and (4). It was clear from those figures that as LSB effect ($L$)
increases, the greybody factor decreases: low values of the greybody factor
indicate a low probability that HR can reach spatial infinity. In the future,
the latter observation might shed light on the LSB effects from both
terrestrial experiments and astrophysical observations. Any discovery of the
LSB would be an important signal beyond the SM physics \cite{Review}.

In future work, we plan to extend the GUP and greybody factor analysis
\cite{Visser} to the various BHs in gravity's rainbow, which is also a result
of quantum gravity
\cite{rainbowv1,rainbowv2,rainbowv3,rainbowv4,rainbowv5,rainbowv6,rainbowv7,rainbowv8,rainbowv9,rainbowv10}%
. The deformation of a spacetime owing to the rainbow gravity effect leads to
Lorentz violations \cite{LVRGv1,LVRGv2}. In this way, we hope to achieve new
results that will help us to understand the QGT and its effect on the LSB.

\section*{Acknowledgements}

The authors are grateful to the editor and anonymous referees for their
valuable comments and suggestions to improve the paper.

\end{document}